\documentclass[comsoc, journal]{IEEEtran}
\IEEEoverridecommandlockouts
\usepackage{cite}
\usepackage{hyperref}
\usepackage{amsmath,amssymb,amsfonts}
\usepackage{graphicx}
\usepackage{textcomp}
\usepackage{xcolor}
\usepackage{soul}
\usepackage{ulem}
\usepackage{algorithm,algorithmic}
\usepackage{textcomp}
\usepackage{booktabs}
\renewcommand{\thetable}{\arabic{table}}
\usepackage{multirow}
 \usepackage{rotating}
 
 \usepackage{array}
  \usepackage{soul}
 \usepackage{lscape, afterpage} 
\usepackage[algo2e]{algorithm2e}
\usepackage{comment}
\usepackage{rotating}

\usepackage{multicol, blindtext}
\usepackage{xcolor}
\usepackage{tikzpagenodes}
\usepackage[a4paper,margin=1in]{geometry}
\usepackage{fancyhdr}
\newcommand{\mari}[1]{\textcolor{black}{#1}}

\begin{document}

\title{Remote Sensing to Control Respiratory Viral Diseases Outbreaks using Internet of Vehicles}

\author{
    Yesin Sahraoui,
    Ahmed Korichi,
    Chaker Abdelaziz Kerrache,\\
    Muhammad Bilal,
    and Marica Amadeo

    \thanks{Y. Sahraoui and A. Korichi are with Kasdi Merbah University of Ouargla, Ouargla, Algeria (emails: sahraoui.yesin@univ-ouargla.dz;ahmed.korichi@gmail.com ).}
    
    \thanks{CA. Kerrache is with University of Ghardaia, Ghardaia, Algeria (emails: kr.abdelaziz@gmail.com;ch.kerrache@univ-ghardaia.dz).}
    
    \thanks{M. Bilal is with Hankuk University of Foreign Studies,
Yongin-si, Gyeonggi-do, Korea (email: m.bilal@ieee.org).}

    \thanks{M. Amadeo is with University Mediterranea of Reggio Calabria, Calabria, Italy (email: marica.amadeo@unirc.it)}
    \thanks{Correspondence should be addressed to  (sahraoui.yesin@univ-ouargla.dz; ch.kerrache@univ-ghardaia.dz)}}

\maketitle
  \begin{tikzpicture}[remember picture,overlay]
    \node[align=center,text=red] at ([yshift=1em]current page text area.north) {Accepted for publication in Transactions on Emerging Telecommunications Technologies (ETT)};
    \node[align=center,text=blue] at ([yshift=-1em]current page text area.south) {\textcopyright  2020 John Wiley \& Sons, Ltd. Personal use is permitted, but republication/redistribution requires  permission.};
  \end{tikzpicture}%

\begin{abstract}

The respiratory viral diseases, such as those caused by the family of coronaviruses, can be extremely contagious and 
spread through saliva droplets generated by coughing, sneezing or breathing. In humans, the most common symptoms of the infection include fever and difficulty in breathing. 
In order to reduce the diffusion of the current “Coronavirus disease 2019 (COVID-19)” pandemic, the Internet of Things (IoT) technologies can play an important role; for instance, they can be effectively used for implementing a real-time patient tracking and warning system at a city scale. Crucial places to install the tracking IoT devices are the public/private vehicles that, augmented with multiple connectivity solutions, can implement the Internet of Vehicles (IoV) paradigm. In such ubiquitous network environment, vehicles are equipped with a variety of sensors, including regular cameras that can be replaced with thermal cameras. Therefore, this paper proposes a new design for widely detecting respiratory viral diseases that leverages IoV to collect real-time body temperature and breathing rate measurements of pedestrians. This information can be used to recognize geographic areas affected by possible COVID-19 cases and to implement proactive preventive strategies that would further limit the spread of the disease.

  
\end{abstract}
\begin{IEEEkeywords}
Internet of Vehicles; Remote Sensing; Viral Diseases; COVID-19.
\end{IEEEkeywords}

\section{Introduction}
\subsection{Background and Challenges}
The new era of the Internet of Things (IoT) innovations enables the smart vehicles to connect to the Internet and to sense and interact with the surroundings, such as pedestrians, roadside units (RSUs) and other smart vehicles. As a result, smart vehicles become the central players of the new Internet of Vehicles (IoV) paradigm \cite{1,abdelaziz2020future}. 

In IoV, the smart vehicles are equipped with sensors, active control systems, GPS receivers, cameras, and embedded storage and processing units \cite{2}. These advanced capabilities not only enable services that enhance safety on the road, help to avoid traffic jam situations, provide entertainment contents, monitor the public healthcare \cite{chen2020secure}, etc., but they also   
materialize the idea of Social Internet of Vehicles (SIoV) \cite{3}. 

SIoV, also known as Vehicular Social Networks (VSN) \cite{4}, is an enhancement of IoV that adds a social dimension to it. In SIoV, the vehicles are capable of establishing social relationships with each other and are allowed to have their own social networks based on their common interests, contexts, goals and intent for socializing \cite{5}.

With the sudden spread of viral diseases like COVID-19 across the world, many lives are under threat; especially older people and those with chronic diseases like cardiovascular disease, diabetes and cancer are at high risk \cite{6}. Thus, detecting the suspected cases has become paramount. This can be accomplished by tracking the most common symptoms of the disease, like body temperature and breathing state, through cameras equipped with thermal sensors. In particular, the idea of using Remote Sensing (RS) data is especially useful in epidemiology, and consists of capturing real-time information about an object or phenomenon, without making physical contact with it \cite{7}. The collected information is then processed for generating knowledge about suspected cases.  

Currently, the family of coronavirus diseases are diagnosed by Polymerase Chain Reaction (PCR) \cite{8}, a method that requires qualified technicians and specialized machines. In addition, the thermal sensors for real-time temperature checks are placed in hot spots, such as seaports, bus terminals, airports and shopping malls.
This approach has the following limitations:

\begin{itemize}
	\item It uses small scale diagnostics and in specific places only.
	\item It is expensive when using PCR techniques; moreover it relies on the availability of \mari{highly specialized} equipment and expertise and takes a lot of time, which reduces the number of diagnosed cases.
	\item It lacks of information about virus spread locations, due to the absence of a unified database.
	
\end{itemize} 
	
Therefore, it is necessary to design an enhanced tool that helps those on the front line to identity and quickly treat the newly infected people.

\subsection{Solution and Contributions}

In this article, we propose a novel respiratory viral diseases detection method, which employs the IoV paradigm.  To effectively identify the infected people and reducing the false positive and false negative results, we detect the suspected cases by considering two factors, namely the breathing rate and the body temperature. In our design, we assume that the vehicles, managed by the police or health departments, are equipped with thermal cameras to detect the body temperature and breathing complications caused by the disease, and with a GPS device to track the location of suspected cases. 
By leveraging these technologies, we collect massive amount of real-time data from pedestrians on the streets and create live statistics in the form of geographic heatmaps. More specifically, the main contributions of this work are summarized as follows:
\begin{itemize}
	\item We propose a new method that leverages IoV to identify, monitor and control viral diseases outbreaks. The conceived method allows to detect possible infections in a very short time at a country level, by covering urban and non-urban areas with accurate localization information. 
	\item Real-time information, collected by vehicles, is delivered to edge computing server(s) where it is temporarily stored, processed and accessed by the interested consumers, e.g., Police, Health Department Inspectors, etc.

	\item We provide a simulation testing of the proposed model in order to evaluate its performance under a variety of metrics, including the time for transferring the sensed information.
\end{itemize}
To the best of our knowledge, the proposed model is the first attempt to utilize IoV and remote sensing for the detection and prediction of respiratory viral diseases.

The rest of this article is organized as follows. We briefly present the related work about IoT-based healthcare, epidemic applications and remote sensing  in Section II. We illustrate the proposed model for remote sensing based on IoV in Section III. How to exploit our model for prevention is proposed in Section IV, while the performance evaluation is reported in Section V. 
Finally, Section VI concludes the work.

\section{Related Work}
This section first presents a description of how IoT is used in healthcare systems, and devote a special focus on the detection of infectious epidemics 
that spread very quickly. Then, we  highlight  representative remote sensing  works that leverage IoV. 
Table \ref{Tab1} compares and highlights the key differences between the considered related work. 

\subsection{IoT Applications in Healthcare}

IoT solutions in healthcare have becoming popular in recent years thanks to their capability of improving the quality of life of patients, by providing the ubiquitous medical services. Healthcare systems enable remote monitoring and real-time diagnosing of diseases, and even remote surgeries over the Internet. Data analysis, detection of critical events and definition of rehabilitation strategies become automated processes, with low human intervention.

\mari{Several works have already discussed the benefits and research perspectives of IoT healthcare systems. For instance,} the authors in \cite{9} give an overview of the IoT applications in the healthcare industry. \mari{First, they identify the enabling communication and sensing technologies and classify the existing smart healthcare devices and applications. Then, they identify the key problems for designing smart healthcare systems, including resource and big data management. The crucial role of social networking in this field is however not investigated.}  

\mari{Similarly,} F. H. Alqahtani et al., in \cite{10}, reviewed the IoT solutions for remote healthcare monitoring systems, and its applications in industry and business, 
and highlighted some issues and challenges. 

S. B. Baker et al., in \cite{11}, identified the key  components  of  an  end-to-end  IoT healthcare system, \mari{and proposed a hierarchical model that includes the following main components. At the user side, \textit{Wearable Sensors} measure the physiological conditions and report them to a close \textit{Central Node}, which processes the sensed information and may implement some simple decision making. Interactions between sensors and the central node are enabled by short-range communication technologies such as Bluetooth. 
Remotely, a \textit{Secure Cloud Storage and Machine Learning} framework collects all the data to make them easily accessible by specialists and nurses and supports advanced processing through artificial intelligence techniques. Long-range communication technologies, such as 3G or Low-Power Wide-Area Networks (LP-WAN), are used to enable the interactions between central nodes and the remote framework.}

The authors in \cite{12} presented the Internet of Health Things (IoHT) concept and summarized the existing services and applications in the field. 
They recognize that the current solutions are mainly growing isolated and lack of interoperability and flexibility. 

In recent years, IoT-based healthcare systems are further improving their diagnosis accuracy thanks to the use of artificial intelligence techniques \cite{gadekallu2020deep,deepa2020ai} and to the deployment of advanced strategies that reduce the energy consumption and costs  \cite{maddikunta2020green}.

Our paper deals with the use of IoT technology to detect viral respiratory infections like COVID-19.  Being this latter a very recent disease, to the best of our knowledge, there are no IoT-based healthcare systems specifically devoted to it. However, some solutions for epidemic detection can be identified in the literature, as explained in the next section.

\subsection{Applications in Epidemics Detection}
o control the spread of contagious diseases, existing works have proposed distinct approaches that range from the use of IoT technologies coupled with cloud computing to fog computing and social sensing. In the following, we discuss some of them.

J. Ginsberg et al., in \cite{13}, presented a detection method based on social sensing for epidemics of seasonal influenza. The proposal relies on the numbers of Google search queries to track influenza-like illness in the population of web search users. However, the limited accuracy of this approach makes it ineffective in practice. 
A similar approach was proposed by C. W. Schmidt in  \cite{14}, where messages posted by people on social media were used to predict and track disease outbreaks.

A. Gupta et al., in \cite{15}, designed a distributed framework for remote health monitoring that leverages Wearable Body Area Networks (WBAN) and sensor-cloud technology to efficiently retrieve data from the body sensors. An algorithm is deployed to minimize the effects of inter-WBAN interference and provide fast, reliable, energy-efficient and fault-tolerant communications. 

S. Sareen et al. in \cite{16}, designed a cloud-based monitoring system for a crowd of patients that uses mobile devices, wireless sensor technology and fog computing for predicting and preventing the Zika Virus outbreak. A GPS system is used to display the infected users with over 90\%  accuracy. 
The same authors in \cite{17} proposed a framework based on Radio Frequency Identification Devices (RFIDs), wearable body sensor technology, and cloud computing to detect and monitor the Ebola virus disease, using the J48 decision tree to categorize infected users.
The lack of an autonomous system to sense data about secondary and advanced symptoms, led however to a decrease of efficiency.

G. Sun et al., in \cite{18}, proposed a combined visible and thermal image processing approach that uses a CMOS camera equipped with infrared thermography to remotely sense multiple vital signs and to rapidly and accurately screen patients who are suspected of carrying infectious diseases. \mari{Based on the measured vital signs, a logistic regression function is used to predict the possibility of infection within 10s. The system has been proved to be effective in presence of small data samples, but its performance in real-world settings has not been evaluated.}

R. Sandhu et al. in \cite{19}, presented an architecture for scanning and controlling the influenza A(H1N1) pandemic using social network analysis and cloud computing. The proposed architecture is tested on a synthetic dataset generated for two million users, but further accuracy and reliability of the predictions are needed, when collecting data from unreliable sources, such as individual users.

A. Mathew et al., in \cite{20}, proposed a smart IoT-based disease surveillance system that relies on 
a main server, located in every hospital, which automatically stores and processes patients' records. If dangerous infections are detected,  the information is immediately transmitted to the Health Ministry though a  central backbone network. By doing so, proper countermeasures can be quickly implemented  to reduce the spreading of  disease. However, the system suffers from high cost, since it requires a dedicated network infrastructure, and it may lack of accuracy, since data are only captured in specific hot spots.

	\begin{table*}[]
	\centering
		 \caption{Comparison of  related work. 
		 }
		 \label{Tab1}
		 		 \rotatebox{90}{
		\begin{tabular}{|p{0.9in}|p{1.0in}|l|l|p{0.9in}|}
			\hline
			\textbf{Reference} & \textbf{Application domain} & \textbf{Advantage} & \textbf{Drawback} & \textbf{Communicaton support}
			\\ \hline
		Baker \textit{et al.}\cite{11} & Healthcare monitoring &  - Comfortable solution for  monitoring the vital signs anywhere, anytime.& - Cloud security threats still the main concerns & WSN, Cloud\\
			& & - Replacement or repair process of wearable nodes is very easy & &

			\\ \hline
		 Ginsberg \textit{et al.} \cite{13}, Schmidt \cite{14}& Epidemic detection  & - Simple method and easy to use & - Suffer from  accuracy problem & Cloud \\
			Sandhu \textit{et al.} \cite{19} &   & & &  \\ \hline
	Gupta {et al.} \cite{15} & Healthcare monitoring & - Reliable, low  cost and secured framework & - The  interference of  WBAN & Cloud, WSN \\ \hline
		Sareen {et al.} \cite{16} & Epidemic detection &  - Provides a high degree of diagnostic accuracy  & - The mobile  phones typically have limited battery life    & IoT, WSN, Cloud, MANET \\
			& & &  - Cloud security issues &  \\ \hline
		Sareen {et al.} \cite{17} & Epidemic detection & - Provides high accuracy in  early symptoms & Consumes a lot of cloud computing resources & IoT, WSN, Cloud \\
			
			& & - Cost-effective solution &  &   \\ \hline
			
		Sun	{et al.} \cite{18} & Epidemic detection & - Rapid screening saves time  &  - Tested only with small samples   & IoT \\ \hline
			
		Mathew {et al.}	\cite{20} & Epidemic detection, & - Fast and accurate solution &  - Suffers from the high cost factor & IoT, Cloud \\ 
		& Healthcare & & & \\  \hline
		Wang and Chen \cite{22} & Air quality monitoring via IoV & - Cost-effective solution & - rely on rewarding policy depending on the destination & WSN, VANET \\ 
			 \hline
			
		Mathur {et al.}	\cite{23} & Monitoring parking space availability  via IoV & - High detection accuracy for parking spaces availability & - Limited battery lifespan for smartphone    & WSN, Cloud, MANET, VANET \\
			  &  & - cost-effective solution &  &  \\ \hline
		Dey	{et al.} \cite{25} & Vehicular traffic application & - Produce traffic updates, and road conditions in real-time & - Detect only few types of vehicles  & WSN, IoT, IoV, Cloud \\ \hline

		\end{tabular}
		}
	\end{table*}
\subsection{Remote Sensing using IoV}
Connected vehicles are provided with a variety of sensors that may gather multiple information about the surrounding environment. So far, the majority of the works have focused on the deployment of smart cities applications targeting traffic control and air quality monitoring \cite{ang2018deployment}. To improve the performance, social interactions have been also considered between vehicles.

For instance, S.C. Hu et al.,  in \cite{21}, proposed a VSN architecture that tracks the concentration of carbon dioxide (CO$_2$) gas remotely, based on GSM short messages and geographic information of vehicles. Data are collected in an edge server, which uses Google Maps to illustrate the result. 
 In the same context, the authors in \cite{22}, leverage remote sensing and VSN to assess the air quality in metropolitan areas.  An efficient data gathering and estimation mechanism is implemented, which demonstrates a significant effectiveness under various scenarios. 
 Vehicle drivers may receive rewards based on their reports sent to the edge server. However, since drivers cannot control the sampling rates of their cars, the architecture needs improved monitoring accuracy.

 S. Mathur et al., in \cite{23}, designed an application named ParkNet, which combines smartphone with ultrasonic range ﬁnders and GPS to track the parking spots availability and report this information to a centralized parking server. The main issue of this architecture is that its performance is affected by the limited power source of the smartphones. 
 
 R. Du \mari{et al.},  in \cite{24}, proposed an effective VSN-based urban traffic monitoring system, which relies on probe vehicles, \mari{e.g., taxis and buses}, and floating cars, \mari{e.g., patrol cars},  for sensing the urban traffic and sending the reports to a traffic-monitoring center. \mari{There, the data are aggregated and analysed to extract meaningful traffic information. The major limitation of this approach is that collected data are related to the roads covered by the considered vehicles and information about specific geographic areas could be missing.}
 
On the same subject, the authors, in \cite{25}, designed an IoV-based traffic monitoring system where a centralised database server collects data from sensors installed on-board of vehicles and provides augmented reality services. However, this architecture needs further enhancements to detect the distinct vehicles types and localize them.  

Unlike previous works, in this paper we leverage the sensing technologies on-board of vehicles to collect body parameters from pedestrians and to infer the presence of a viral respiratory infection. To improve the accuracy of data gathering, our system also leverages the social relationships between vehicles.

\section{Proposed Model}

By combining  sensing process and networking, IoV represents a key and cost-effective technology for accessing user data, including epidemic information. 
In this section, we present the proposed framework that leverages IoV for the quick detection of viral diseases through the sensing of body temperature and respiratory rate of pedestrians. The considered parameters may be symptoms of several viral infections, including seasonal influenza. However, since today COVID-19 is highly prevalent worldwide and poses a serious threat to citizens' health, in the following we specifically refer to it.

\subsection{Motivations and Basics}

It is worth to observe that many people affected by COVID-19 could have mild or no symptoms at all, and therefore the viral infection cannot be detected when measuring body temperature and breathing rate.
At the same time, however, previous studies on influenza, and more recent studies on COVID-19, are demonstrating that the viral loads in asymptomatic carriers are relatively low, see \cite{yuen2020sars} \cite{he2020relative} and references therein. Therefore, the relatively transmissibility of asymptomatic cases could be significantly smaller than that of the symptomatic cases. \\
Our system is designed to detect symptomatic cases that are the most dangerous ones. Although people are recommended to stay at home when showing influenza-like symptoms, multiple events have been reported about people affected by COVID-19 that violated quarantine, even with high fever. Sometimes, they were accidentally recognized by the police in railway or bus stations.
Therefore, by supporting the autonomous recognition of symptomatic patients, the proposed system may have a crucial role in fighting the pandemic.

Our design is based on two main actors: vehicles and edge computing servers. The data collected by vehicles in a geographic area are forwarded to an edge computing server where they are temporarily stored, processed and accessed by the interested consumers. According to the 3GPP-V2X (vehicle-to-everything) specification in \cite{23-285}, a V2X Application Server can be implemented according to the Multi-Access Edge Computing (MEC) paradigm to support V2X applications. By providing storage and computing resources close to where data are produced, the V2X Application Server ensures that data are processed in real-time at the network edge thus also limiting the traffic load in the core network \cite{amadeo2019enhancing}. 

In addition to traditional road traffic applications, the V2X Application Server can be extended to support a variety of IoV applications, including those for epidemic detection.
Of course, multiple V2X Application Servers can be deployed in different geographic areas and their storage and processing resources can be sized according to the population density. Depending on their role, the interested consumers can access the data from a single edge server or from multiple ones. For instance, Health Department Inspectors working on a specific area will access only the data from that area; vice versa, if the Health Ministry is interested in an overall map of the suspected infection cases, the data from all the V2X Application Servers will be accessed.

\subsection{Remote Sensing and Processing Operations}
As shown in Figure 1, the proposed framework consists of two sections: a back-end section including the edge server, and a front-end section including vehicles equipped with GPS and thermal camera sensors. The workflow related to the remote sensing tasks for infection detection is instead depicted in Figure 2. 
It starts with the data acquisition and ends with the decision making about whether an intervention is needed to contain the spread of the infection. More specifically, the following stages are considered.

\begin{figure}[htbp]
	\centerline{\includegraphics[width=1\columnwidth]{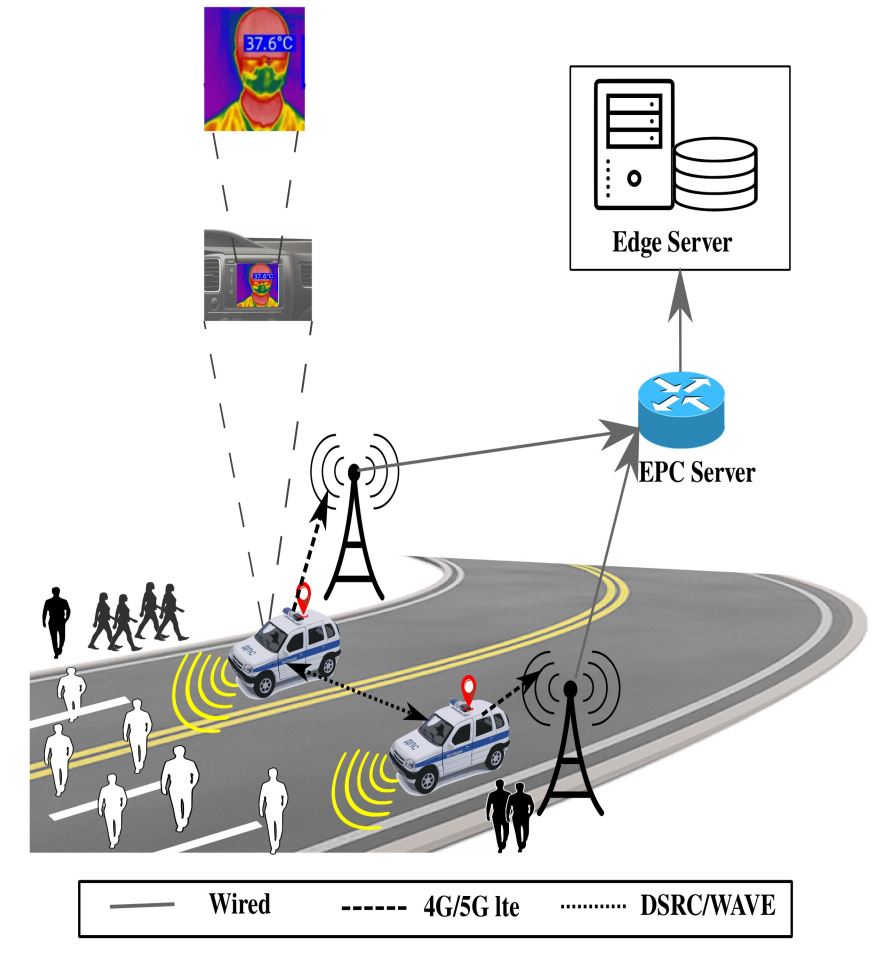}}
	\caption{The proposed model for identifying suspected cases of the COVID-19 virus.}
	\label{fig}
\end{figure}

\begin{figure}[htbp]
\centerline{\includegraphics[width=1\columnwidth]{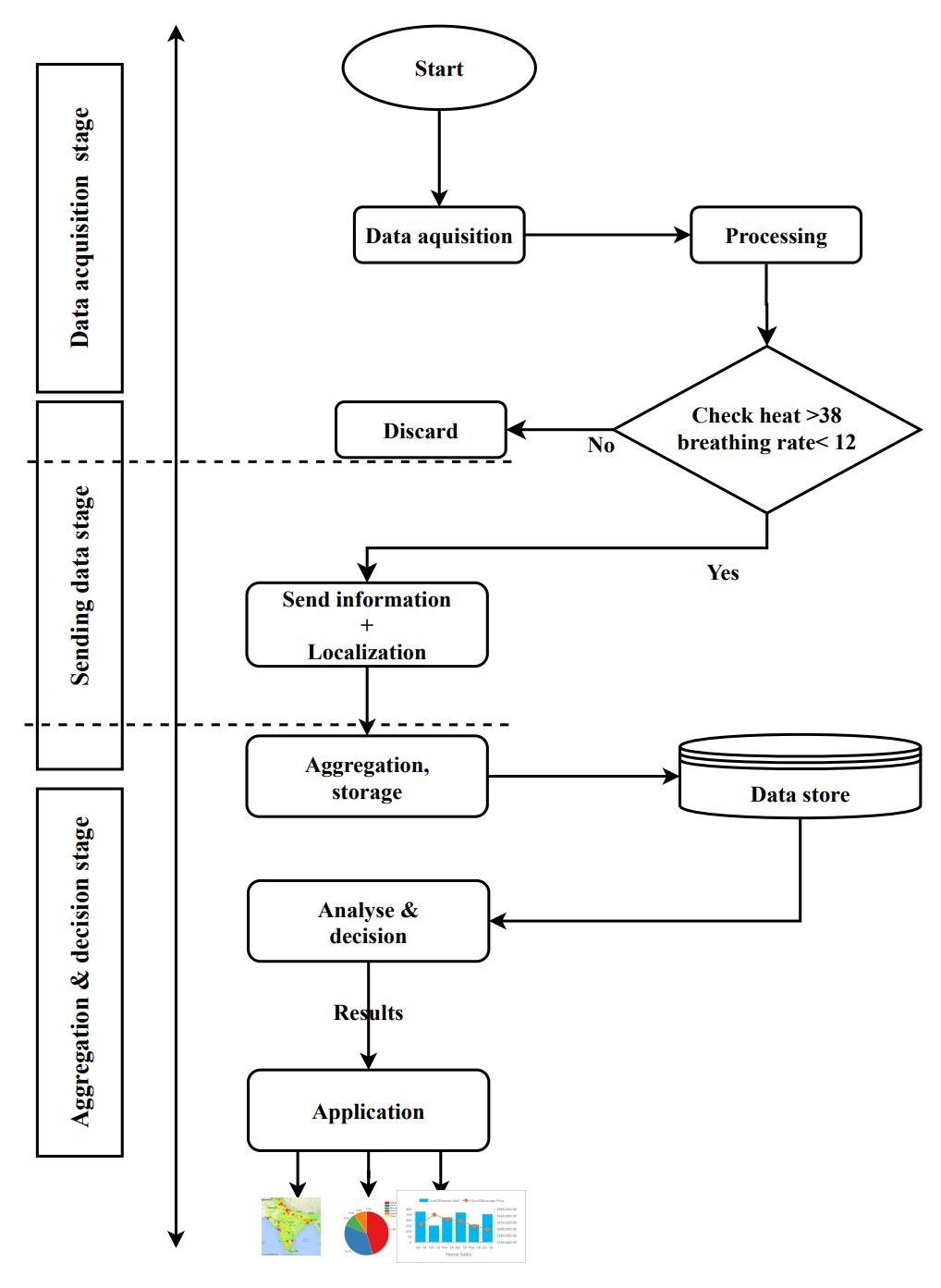}}
\caption{Flowchart of the main stages for our proposed model.}
\label{fig}
\end{figure}  

\begin{itemize}
	\item \textbf{Data acquisition stage.} The first step in the process of detecting suspected cases is the data acquisition, which leverages  Vehicle to Pedestrian (V2P) communications.
	Emergency and patrol vehicles, like police cars and ambulances, are equipped with thermal cameras for checking the body temperature and calculating the breathing rate of pedestrians on the street. 
	More specifically, each thermal camera is set for scanning pedestrians' bodies on both sides of the streets, but its movement can also be controlled by the operator (e.g., policemen and ambulance men) in order to reach the narrow corners and to sweep the wide areas.
    By monitoring how the temperature of the nasal area changes during inhalation and exhalation, the system is able to calculate the breathing rate and to infer if a person is affected by shortness of breath \cite{26}.
    
    After the data acquisition, a local processing is performed to detect possible infections from the measured data.
	It consists of checking if the minimum body temperature rises above the normal threshold\footnote{The normal body temperature or normothermia ranges from 36.5\textcelsius to 37.5\textcelsius \cite{27}.} and if there are  respiratory complications \cite{28}.
	Data related to people that show no symptoms can be simply discarded.
	Conversely, in case potential infections are detected, the information is combined with the geographical coordinates of the pedestrians by using GPS localization.  
   
In order to reduce the data collection time and avoid duplicated data, the vehicles work in a collaborative environment: each one is responsible to sense the suspected cases in some predetermined city streets. 
Our model also leverages the social dimension of smart vehicles to let them exchange information about their specific covered (and not covered) areas in order to: \textit{(i)} avoid redundant measurements in sensing tasks, and \textit{(ii)} to share crucial important information, such as places of mass gathering, which need assistance and reinforcement for rapid screening of infections.

	\item \textbf{Sending data stage.} \mari{It includes a two-step transmission process to relay the information about suspected cases and their position to the edge server.} First, vehicles transmit real-time information 
	to the \mari{closed} Road Side Unit (RSU), using Vehicle-to-Road Infrastructure (V2I) communications. Then, the RSU leverages Vehicle-to-Broadband Cloud (V2B) communication \cite{29}, e.g., by exploiting 
	4G/5G LTE technology to cover both urban and rural roads, and  forwards the information to the \mari{edge server, as shown in Figure 1.} 

	\item \textbf{Aggregation and decision stage.} It includes the process of aggregating, storing and analysing in the edge server all the data generated by the sensors in real time. The output of this stage includes the creation of geographic heatmap reports, which can be sent to the Health Ministry for a prompt assessment. This allows to make decisions that can prevent or slow the spread of infections like COVID-19, including putting people in quarantine or under medical surveillance, performing viral tests and disseminating alert announcements, e.g., through the use of online social networks (OSN) or SMS.  

\end{itemize}

\section{OSN/SMS based prevention}
OSNs are a fast avenue to create and spread announcements about possible risks of infections in specific geographical areas \cite{30}.
As the use of smartphones with social applications  increases in societies, information within popular OSN can have a great impact on prevention support. Therefore, we can exploit our framework to promptly share the information collected in the edge server via OSN. Of course, to guarantee the reliability and credibility of the information, the dissemination must be performed through the use of official social media accounts, such as those of the Health Ministry.


Another great and cheap way to notify people of possible exposure to the infection consists in sending free short messages (SMS). Again, to ensure the credibility of the information, the operation should be performed by the Health Ministry in coordination with all Mobile Network Operators (MNO). 
Thanks to the geo-location information provided by the smartphones, it will be possible to alert only people in the affected regions. The notification may also include multiple suggestions, such as to limit social interactions, to wear a face mask, etc. 
It is worth noticing that the conceived strategy does not compromise the privacy of pedestrian users. Indeed, the collected information is completely anonymous and only the location of possible infection cases is notified.

Figure 3 summarizes how to exploit the proposed model for prevention purposes.

\begin{figure}[htbp]
	\centerline{\includegraphics[width=1\columnwidth]{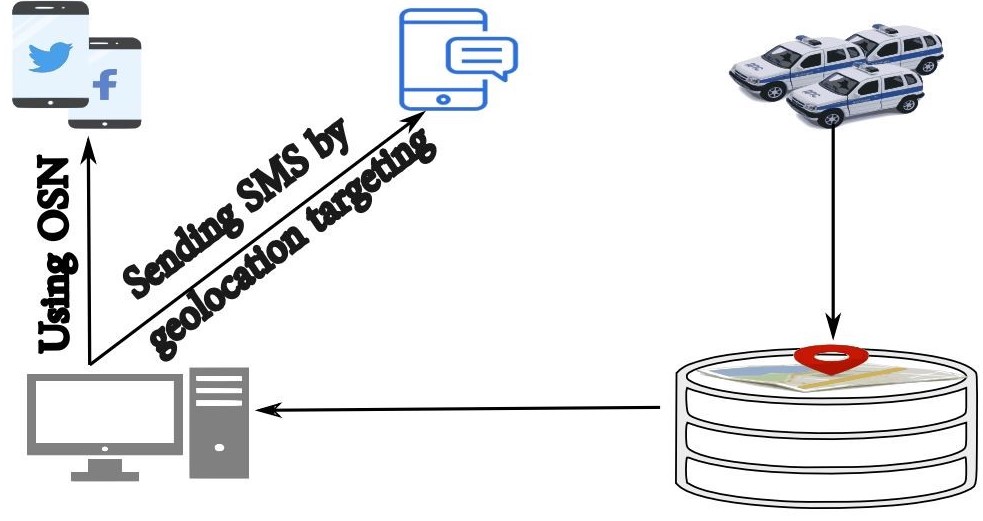}}
	\caption{Exploiting the proposed model for prevention.}
	\label{fig}
\end{figure}

\section{Performance evaluation}
\subsection{Synthetic Data Generation}
To perform a quantitative analysis of the proposed system, we need a huge set of input data about COVID-19 infection cases.  Therefore, we created a synthetic dataset that generates suspected cases
according to the number of COVID-19 cases that are daily reported in specific geographic areas and the correspondent population density. Through this information, we defined an infection threshold $T$, equal to the ratio between the discovered infection cases and the inhabitants in that zone. For instance, if the population in a certain area is about $P=1000$ people and, there,  $C=10$ cases of coronavirus have been discovered so far, we compute the threshold as  $T=\frac{C}{P}=0.01$. When building the synthetic database for that specific geographic region, we consider that threshold to simulate the ratio of infected pedestrians.

As shown 
in the pseudocode of Algorithm 1, we consider a population of users and randomly assign values of body temperature and breathing rate to them.  
When the temperature exceeds 38\textcelsius{}, and the breathing rate is slower than 12 breaths per minute \cite{27},  
then the user is considered as a new potential infected case.
 
 We took Annaba city, Algeria, as a sample, and we divided it into five regions according to the population density. Table 2 shows the number of suspected cases vs. the number of pedestrian users per each zone, obtained according to Algorithm 1.
 
 	\begin{algorithm}
		\caption{Generation of synthetic dataset}
		\SetKwInOut{Input}{Input}
		\SetKwInOut{Output}{Output}
		\Input{P a number of pedestrians at a particular location and C a number of reported  cases per P population.}
		\Output{Number of suspected cases in this location.}
		 \textit{Let s be the number of suspected cases initialized with 0},\\
		 \textit{Let T be the infection threshold, where T $\gets$ $\frac{C}{P}$}:\\
		\While{$\frac{s}{P} \leq T$}{
			$\text{  Assign random value to fever symptom} \\
			\text{ Assign random value to breathing rate}$ \\
			\eIf {(the fever generated value $>$ 38 \textcelsius{})  \textbf{and}  \\(the breathing rate generated value $<$ 12 rpm)}
			{ Add as suspected case.\\
			Increment $s$ by one.
	     	}{
		     Discard the new case.
			 }
		}
	\end{algorithm}

\begin{table}[htbp]
	\caption{Synthetic dataset of suspected cases.}
	\begin{center}
		\begin{tabular}{|c|c|c|}

			 \cline{2-3}
		
			  \multicolumn{1}{c|}{}  & \textbf{Pedestrians}	 & \textbf{suspected cases} \\ \hline
			 
			zone 1 &	38000 &	6500 \\  \hline
			zone 2 &	24000 &	1200 \\ \hline
			zone 3 &	12000 &	4500 \\ \hline
			zone 4 &	5000  &	2000 \\ \hline
			zone 5 &	4000  &	700  \\ \hline

			\hline
			
		\end{tabular}
		\label{tab1}
	\end{center}
\end{table}

Such data are used to create a geographic heatmap representing the density of potential infections in different areas of the Annaba city in Algeria. As depicted in Figure 4, by displaying the density of suspected cases with geographic points as a gradient color layer to style the map, we can create an early alerting and reporting project.

\begin{figure}[htbp]
	\centerline{\includegraphics[width=1\columnwidth]{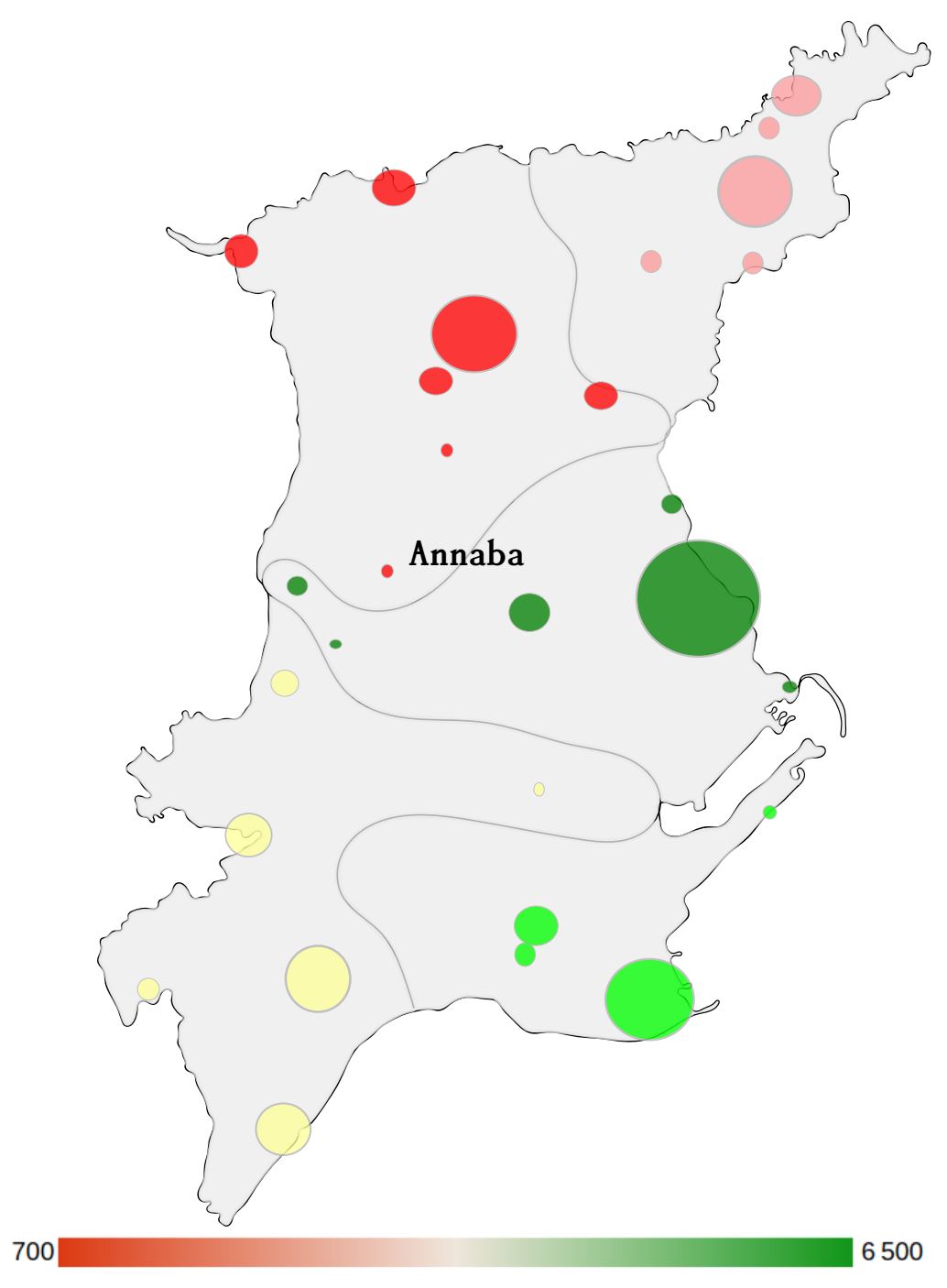}}
	\caption{Suspected cases of viral diseases heatmap.}
	\label{fig}
\end{figure}

\subsection{Simulation and Results}
The performance of the proposed model was evaluated 
by using the network simulator-3 (ns-3) \cite{31}, an open source discrete-event based network simulating software, developed for research and educational purposes. 

In our scenario, the vehicles' routes and mobility model of Annaba city are generated using the Simulation of Urban MObility (SUMO) \cite{32} tool and the  output file is imported into ns-3. The synthetic dataset of the suspected cases is given as input to the vehicles, which send the information to the edge server. We assume that each data packet carries localization information, body temperature and breathing rate of only one case.
Vehicles leverages the LTE technology to connect to the edge server, as foreseen in the proposed model. 
More specifically, in addition to the LTE radio access network, the simulation scenario includes the Evolved Packet Core (EPC) network, as shown in Figure 1, which connects to the edge server via a point-to-point link.

\subsubsection{Simulation Setup}
General settings and LTE network parameters taken into account in the experimentation are summarized in Table 3. 

In order to measure the performance of the model when varying the vehicles density, we consider a number of vehicles ranging from 2 to 40 for each geographical zone. 
Non-line-of-sight (NLOS) channel conditions are appropriately modeled by using the Nakagami-m distribution. 

\begin{table*}[ht]
	\centering
	\caption{Simulation parameters.}
	
	\begin{tabular}[t]{ccc}
		\toprule

		&\textbf{Simulation parameters}  &   \textbf{Values}\\
		\midrule
		
		\multirow{4}{*}{\rotatebox[]{90}{\textbf{General}}}&Communication technology &  LTE + wired\\
		&Simulation Time& 50 seconds\\
		&Vehicle density & 2, 6, 10, 20, 40\\
		&Number of eNB& 2\\
		&Number of pedestrians&4000, 5000, 12000, 24000, 38000\\
		&Ratio of contaminated pedestrians& 17.5\%, 40\%, 37.5\%, 5\%, 17.1\%\\
		&Data packet size&1024\\ \hline
		\multirow{4}{*}{\rotatebox[]{90}{\textbf{LTE}}}&Propagation loss model &Nakagami\\		
		&LTE data packet type & TCP\\		
		 &Data rate/RB allocation & DL (50)$/$UL (50) \\
		  &Transmission power & eNB (49 dBm)$/$UE (23 dBm) \\
		 
		\bottomrule
	\end{tabular}
\end{table*}%

\subsubsection{Performance Metrics}
To evaluate the performance of the proposed model, we consider the following network metrics. \\

\textbullet \quad End-to-end delay (E2E delay). It is defined as the time taken by a packet to be transmitted across the network from the vehicle (sender) to the edge server (destination). It is computed as the average  ratio between the sums of all delays and the total number of received packets, as follows:
	
\begin{gather}
Mean\_E2E\_delay =
\frac{\sum_{}delaySum_{}}
{\sum_{}receivedPackets},
\intertext{where:}
\begin{tabular}{>{$}r<{$}@{\ \ }l}
delaySum = \sum\limits_{\forall j \in received Packets} received_j-sent_j.
\end{tabular}\nonumber
\end{gather}

Figure 5 shows the average end-to-end delay when varying the number vehicles in each zone. It can be observed that, reasonably, the delay increases with the number of vehicles, i.e., packet congestion occurs when a large number of vehicles send the packets to the destinations over the resource-limited wireless channel. 
However, even in the higher density settings (40 vehicles), the delay is always less than 1 second, thus guaranteeing that information is collected in a timely fashion. 

\begin{figure}[htbp]
	\centerline{\includegraphics[width=1\columnwidth]{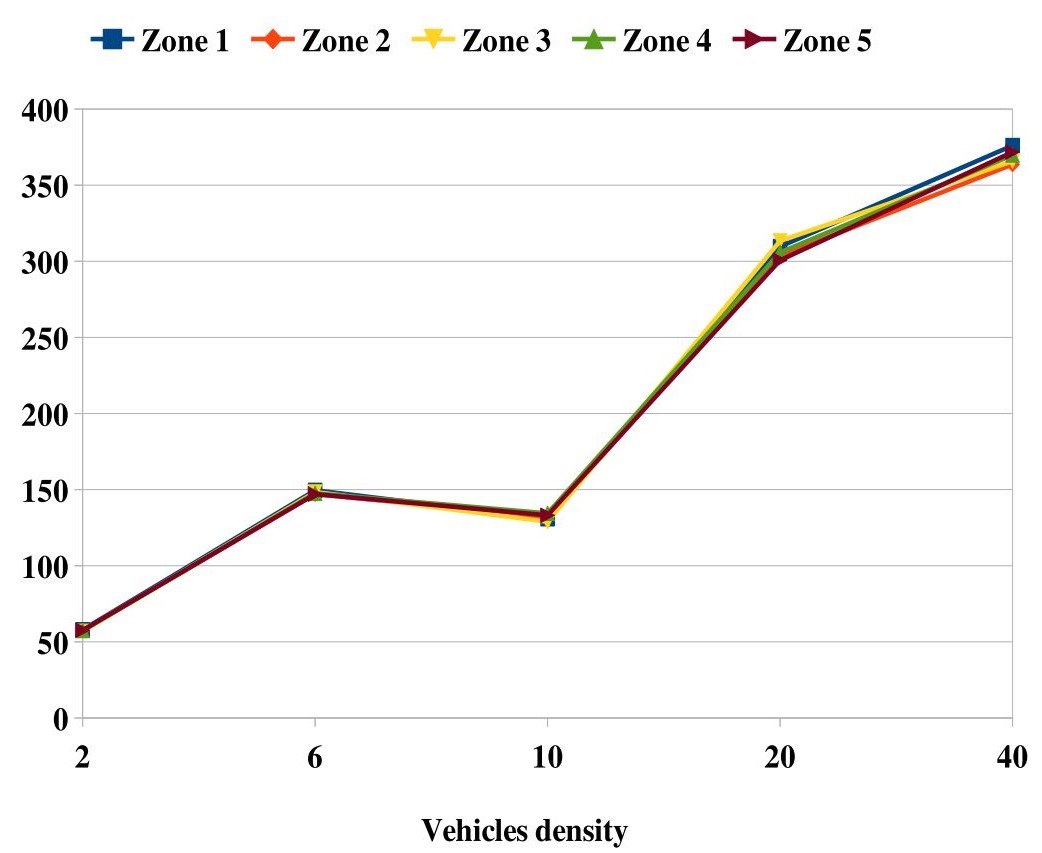}}
	\caption{Mean E2E delay variation according to vehicles density.}
	\label{fig}
\end{figure}

\textbullet \quad Packet Delivery Ratio (PDR). It is defined as the ratio between the number of packets received by the edge server, and the number of packets transmitted by vehicles, during the simulation time, see Eq. (2).
	
	\begin{gather}
PDR =
\frac{\sum_{}receivedPackets_{}}
{\sum_{}transmittedPackets_{}},
\end{gather}

As shown in Figure 6, it can be observed that our model is able to maintain a  relatively high packet delivery ratio, even when the vehicles density increases. 

\begin{figure}[htbp]
	\centerline{\includegraphics[width=1\columnwidth]{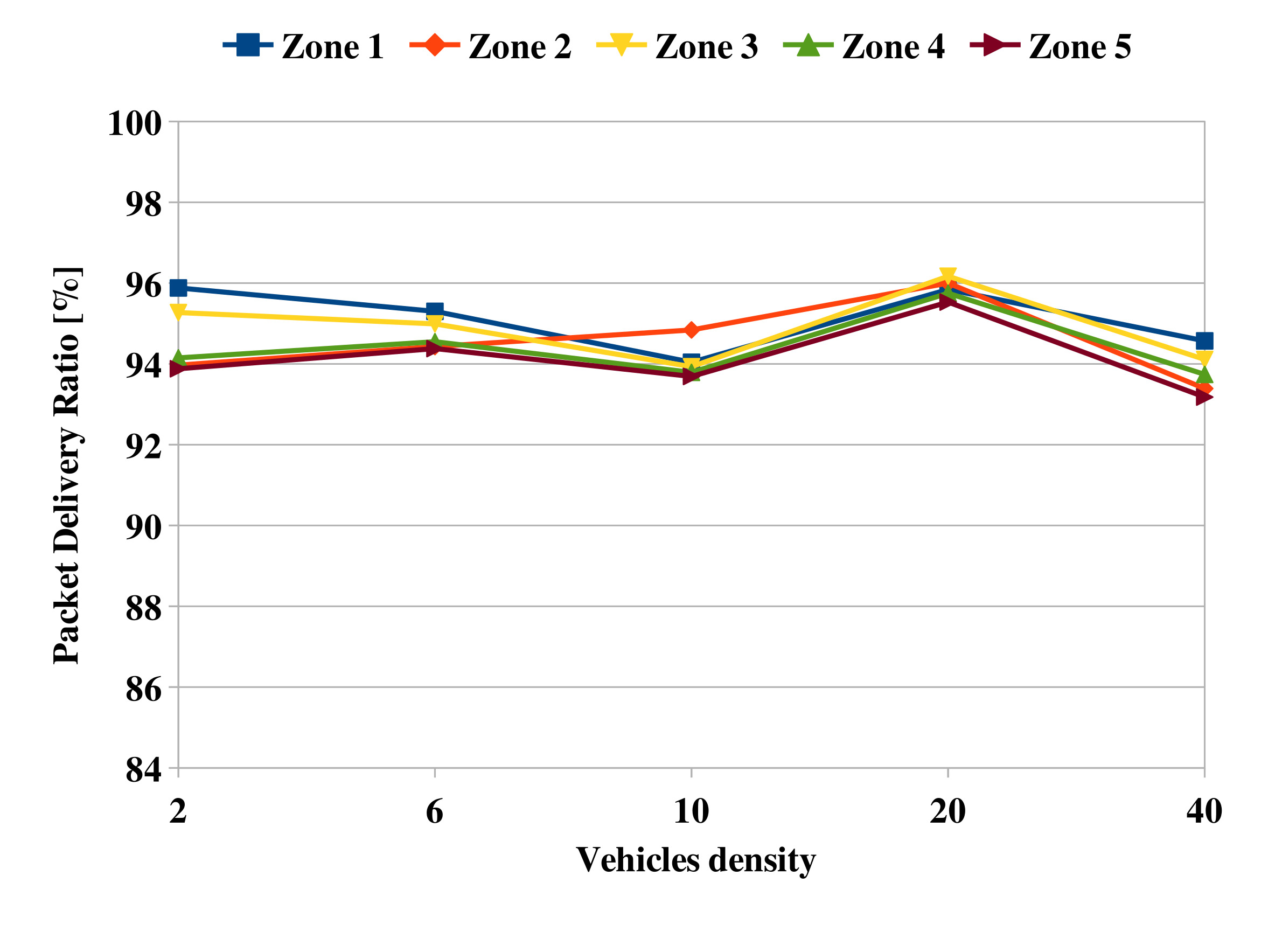}}
	\caption{Average PDR variation according to vehicles density.}
	\label{fig}
\end{figure}

\textbullet \quad Packet Loss Ratio (PLR). It represents the ratio of lost packets to the total number of sent packets, see Eq. (3).

\begin{gather}
PLR =
\frac{\sum_{}lostPackets_{}}
{\sum_{}transmittedPackets_{}}= 1- PDR,
\end{gather}

As shown in Figure 7, the PLR is used to evaluate the loss performance of our proposed model.  It can be observed that the PLR is inversely proportional to PDR.  Indeed, when multiple competitors share the same limited channel resource, it is more likely that packets are lost due to collisions.
Whenever this factor is small, the framework is suitable to send real-time sensing data. 

\begin{figure}[htbp]
	\centerline{\includegraphics[width=9cm,height=6cm]{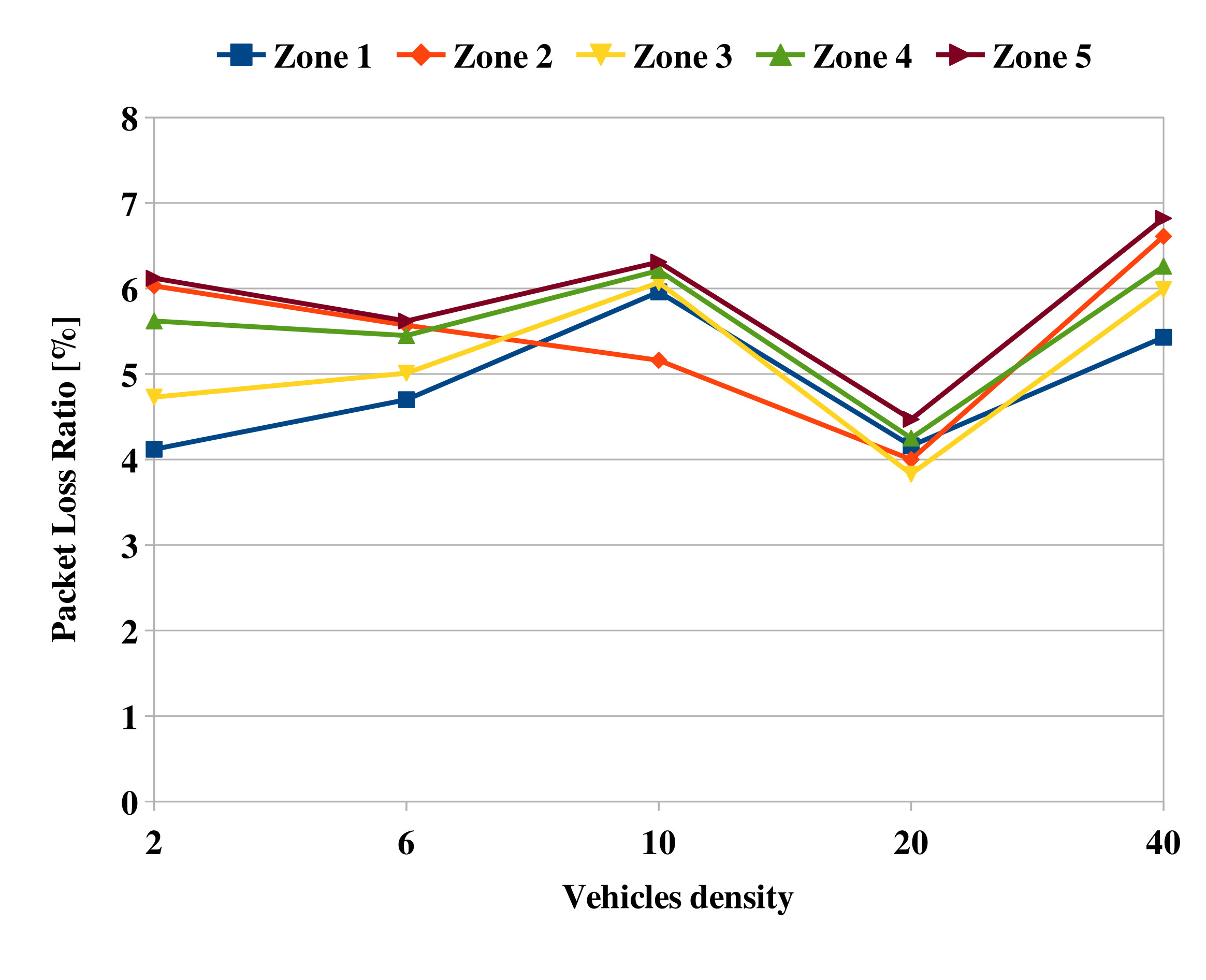}}
\caption{Average PLR variation according to vehicles density.}
\label{fig}
\end{figure}

\textbullet \quad Throughput. It represents the amount of data successfully received by the edge server per unit time, measured in kilobits per second (kbps), as shown in Eq. (4).

\begin{gather}
Throughput=
\frac{\sum_{}received\_ data_{(kbps)}}
{T_{last}-T_{first}},
\intertext{Where:}
\begin{tabular}{>{$}r<{$}@{\ :\ }l}
T_{last} & is the time of last packet send. \\
T_{first} & is the time of first packet send.\\
\end{tabular}\nonumber
\end{gather}

Figure 8 shows that, reasonably, the throughput increases with the vehicle density  because a higher number of packets are received by the edge server. 

\begin{figure}[htbp]
	
	\centerline{\includegraphics[width=1\columnwidth]{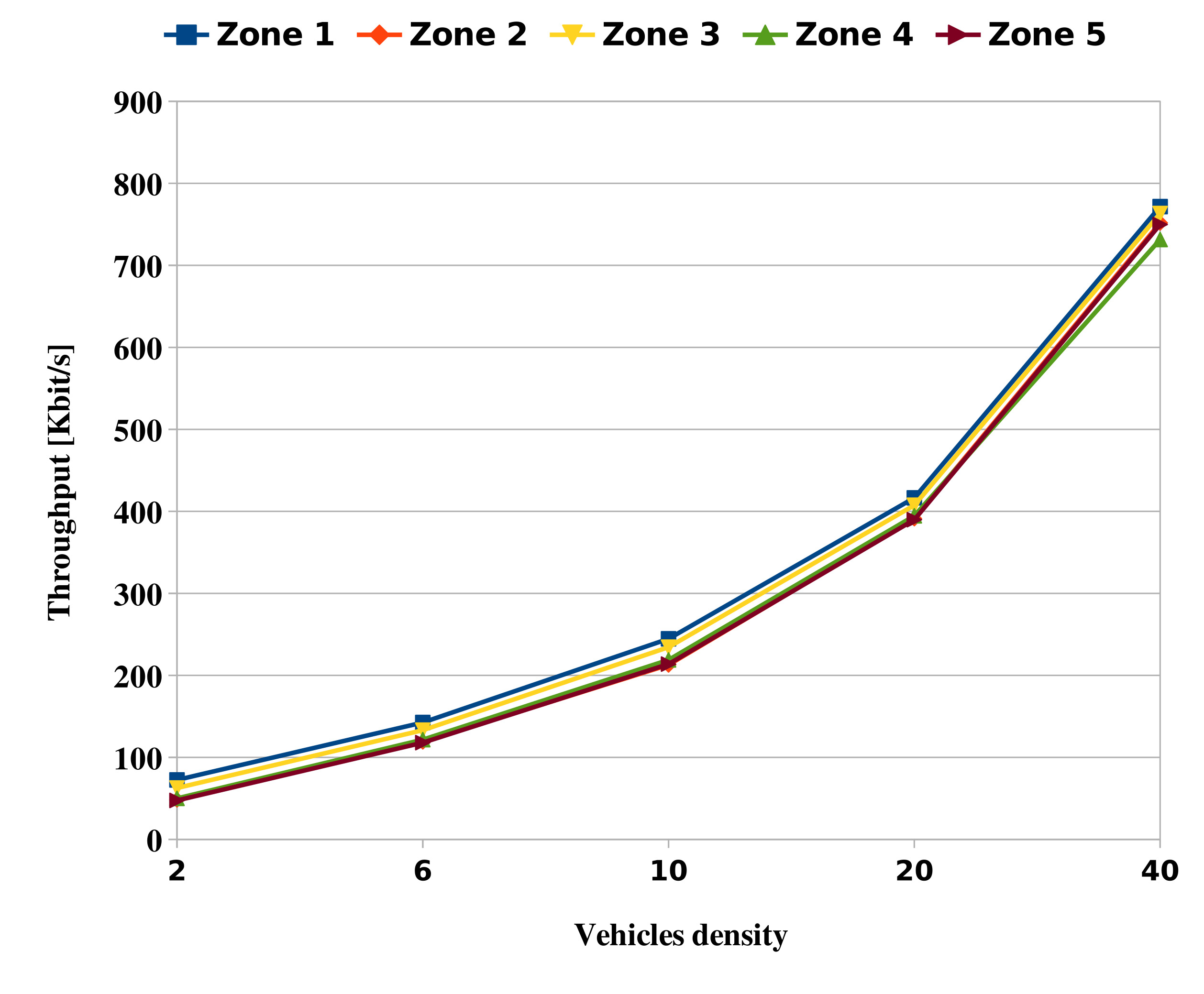}}
	\caption{Throughput variation according to vehicles density.}
	\label{fig}
\end{figure}

\begin{figure}[htbp]
	\centerline{\includegraphics[width=1\columnwidth]{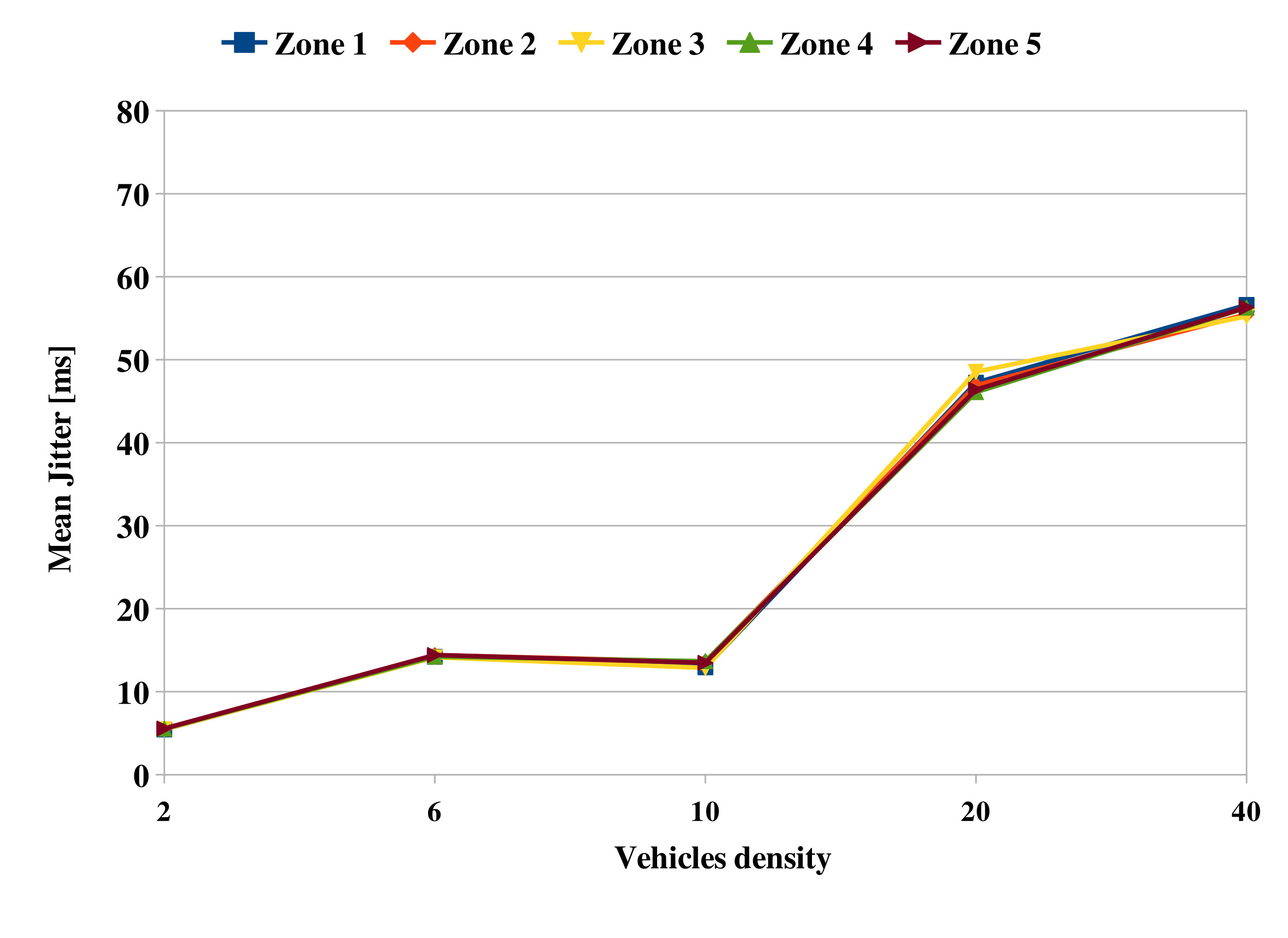}}
	\caption{Mean delay jitter in respect of vehicles density.}
	\label{fig:jitter}
\end{figure}

\begin{itemize}
	\item {Delay Variation (or Delay Jitter):} jitter is defined as the variation in the end to end delay of packets, caused by some factors such as congestion. The performance increases as jitter decreases.
	
	The mean jitter can be calculated using the formula given in Eq. (5):
	
	\begin{gather}
Mean\_jitter =
\frac{\sum_{}jitter_{}}
{\sum_{}receivedPackets_{}-1},
\end{gather}

Figure \ref{fig:jitter} shows that the delay jitter increases with the vehicle density,  due to the higher congestion. However, in the worst case, the metric is below 60 ms and therefore the system performance is not compromised. 
\end{itemize}

\subsection{Accuracy measures}
It is well known that remote sensing can be affected by several factors, e.g., people could show mild symptoms of the disease or they could develop fever and respiratory symptoms simply due to a common cold. This can lead, respectively, to the occurrence of false negative or false positive results:

\begin{itemize}
	\item\textbf{False negatives (FN):} A test result which falsely indicates the absence of a condition or an outcome. 	
	\item\textbf{False positives (FP):} A test result which falsely indicates the presence of a condition or an outcome. 
\end{itemize}

To test the accuracy of the proposed model, we consider the zone 4 as an example. As shown in Figure 10, we assume that, for the total tested population, there is a false positive rate of 4\% and a false negative rate of 1\%; these numbers are in line with those reported in existing literature about remote sensing through thermal imagining cameras \cite{33,34,35}.

\begin{figure}[htbp]
	\centerline{\includegraphics[width=1\columnwidth]{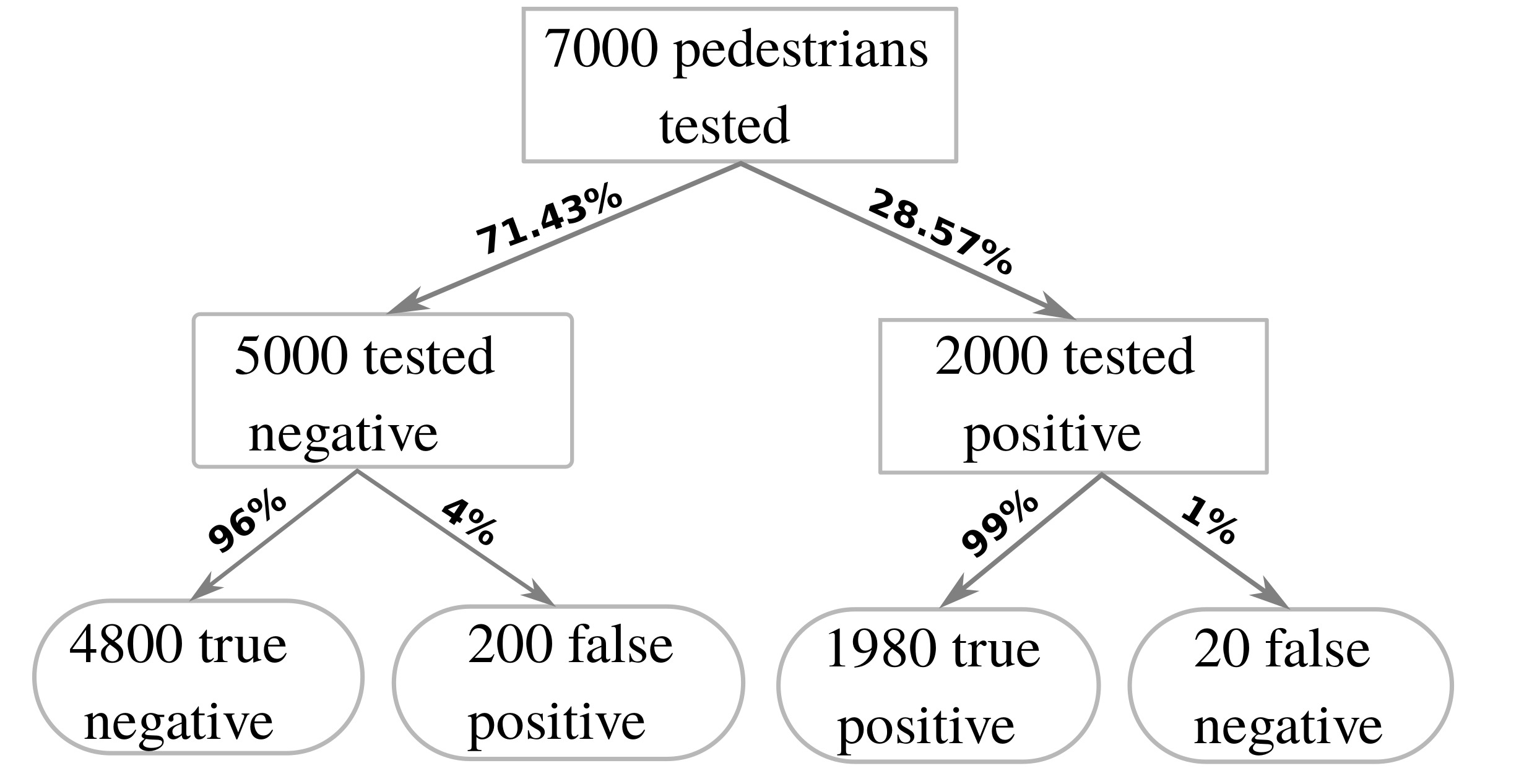}}
	\caption{Tree diagram \mari{for pedestrians tested in zone 4}.}
	\label{fig}
\end{figure}

In our model, we identify possible infections when the temperature is absolutely greater than 38\textcelsius, and the breathing rate is less than 12 rpm. 
If pedestrians tested positives, they require treatment, isolation and quarantine, thereby preventing the infection of others. Of course, having false positives is better than having false negatives, which lead to bad outcomes on the public health.

The graph in Figure 11 shows the false positive (FP) and false negative (FN) results around 38\textcelsius. Similarly, for the breathing rate, we take false positive (FP) and false negative (FN) results around the threshold, where the threshold is 12 rmp. 

\begin{figure}[htbp]
		\centerline{\includegraphics[width=1.1\columnwidth]{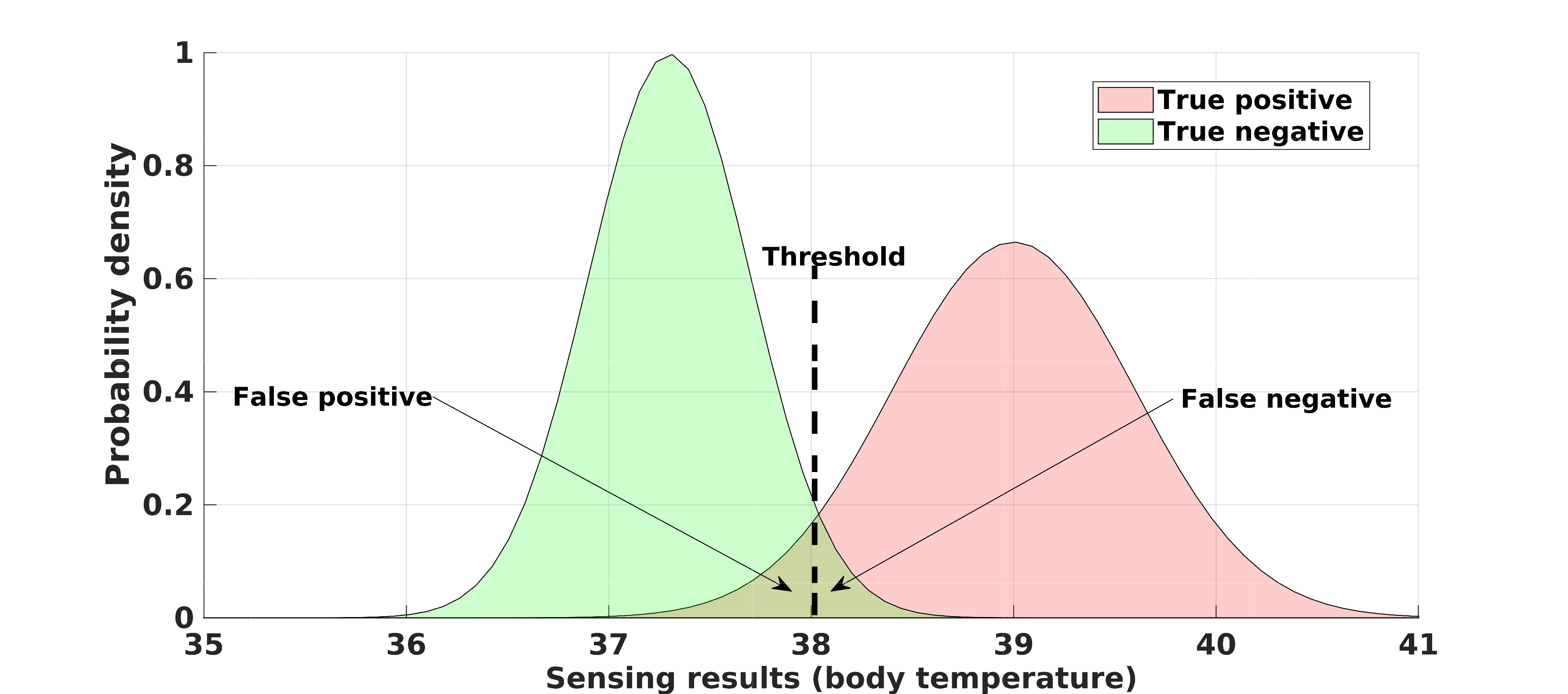}}
	\caption{Graph to represent false negative and false positive.}
	\label{fig}
\end{figure}
From the tree diagram, we have:

\begin{gather}
FN = P(D \cap \overline{P})= 80,
\end{gather}

\begin{gather}
FP = P(\overline{D} \cap P) = 50,
\intertext{Where:}
\begin{tabular}{>{$}r<{$}@{\ :\ }l}
P & the sensing result is positive. \\
D & the pedestrian has the disease.\\
\end{tabular}\nonumber
\end{gather}

 In addition to false negatives and false positives, other most common accuracy measures are sensitivity and specificity.
 \begin{itemize}
  \item \textbf{Sensitivity:} also referred as the \textit{recall}  or True Positive Rate (TPR), it represents the ability of diagnoses procedure to correctly identify all patients with a positive test (true positives). A very sensitive test is able to correctly identify all positive samples. From the tree diagram we have:
 \end{itemize}

\begin{gather}
Sensitivity =
\frac{TP}
{TP + FN} = P(P\mid D),
\end{gather}
\begin{itemize}
	\item \textbf{Specificity:}  also referred as the  True Negative Rate (TNR), it represents the ability of a diagnoses procedure to correctly identify all patients with a negative test (true negatives). Sensitivity and specificity are inversely proportional, if the first increases, the second decreases and vice versa. A good diagnostic test which identifies all false positives and false negatives, consists of  a test with high sensitivity and low specificity combined with another test having high specificity and low sensitivity. From the tree diagram we have:
\end{itemize}

\begin{gather}
Specificity =
\frac{TN}
{TN + FP} = P(\overline{P}\mid \overline{D}),
\end{gather}

It is worth observing that the accuracy of the results in our design is not 100\% perfect, but this is in-line with other diagnostic tests.

 In presence of infectious diseases, false negative tests may have dangerous implications for the public health. 
Similarly, false positive tests poses different threats. For instance, the already overburdened medical facilities are used for patients that are actually healthy, or false positive diagnosed persons may experience adverse psychological consequences until the condition is not clarified.

To improve the accuracy of data collection, additional detecting strategies can be included in our system, e.g., based on artificial intelligence modelling techniques \cite{33,34,35}. However, this aspect is left as future work.



\section{Discussion and Conclusion}



Today, COVID-19 is one of the most prevalent diseases in the world, taking thousands of lives every day. In this article, we presented an IoV based remote sensing framework to identify and control the outbreaks of infectious disease like COVID-19.
 The proposed system is  designed to scan the body  of a huge number of pedestrians rapidly, by using thermographic cameras, and  to transmit the data to  the edge server in real time. The GPS is used to track the infected areas. Finally, to quickly spread the information about risk-prone areas and help the people to protect themselves, the use of OSN/SMS is proposed.
Sensed data can be further transferred  in a unified database, thus  building a smart city sensing system.

A main challenge to our model is the deployment cost due to the installation of thermal cameras on board of vehicles, which may not be an affordable option for underdeveloped countries. 
However, the deployment cost can be highly rewarded by the benefits of the system, as showed by our preliminary evaluation campaign measuring different QoS metrics.
Results demonstrated that our proposal is able to collect and process in real-time thousands of data and the performance is directly affected by the vehicles density. To further reduce the packet losses and mitigate the network congestion, additional assessments with an increased number of eNBs will be considered.

Future works will also include the evaluation of security risks and the design of strategies to protect the integrity of data. 




\bibliographystyle{IEEEtran}

\bibliography{mybibfile}

\end{document}